\documentclass[twocolumn,showpacs,preprintnumbers,amsmath,amssymb]{revtex4}

\usepackage{graphicx}
\usepackage{dcolumn}
\usepackage{bm}

\begin{document}

\preprint{cond-mat/0701000}

\title{FPU phenomenon for generic initial data}

\author{A.~Carati}
 \email{carati@mat.unimi.it}
\author{L.~Galgani}%
 \email{galgani@mat.unimi.it}
\author{A.~Giorgilli}
 \email{giorgilli@mat.unimi.it}
\author{S.~Paleari}
 \email{paleari@mat.unimi.it}
\affiliation{Department of Mathematics,  University of Milano \\
                 Via Saldini 50, I--20133 Milano, Italy.}

\date{\today}

\begin{abstract}
The well known FPU phenomenon  (lack of attainment of 
equipartition of the mode--energies
at low energies, for some  exceptional initial data) 
suggests that the FPU model does not have the
mixing property at low energies. 
We give numerical indications that this is actually
the case. This we show by computing orbits for sets of initial data of
full measure, sampled out from the microcanonical ensemble 
by  standard Montecarlo techniques. Mixing is tested by looking at the
decay of the autocorrelations of the mode--energies, and it is found
that   the high--frequency modes have  autocorrelations that tend instead 
to positive values. 
 Indications are given that 
such a nonmixing property  survives in the thermodynamic limit. It is
left as an open problem whether mixing obtains within time--scales
much longer than the presently available ones.
\end{abstract}

\pacs{Valid PACS appear here}
\maketitle

By the ``standard'' FPU phenomenon we mean the celebrated one observed
for the first time in the year 1955 \cite{fpu}.  
Namely, in numerical  integrations of the
equations of motion for a chain of  particles coupled by weakly
nonlinear springs,
equilibrium is not attained within the available  computational time, 
and a kind of
anomalous  pseudoequilibrium does instead show up. This is observed at
low energies, for
initial data  very far from equilibrium. The quantities studied  
were the energies $E_j(t)$ of the
normal modes  of the linearized system, and their time--averages
$\overline{E_j}(t)$  were found to  relax each to a different value rather
than to a common one, against the  equipartition 
principle (see especially the last figure of the original FPU report).

It was later found by Izrailev and Chirikov \cite{iz-ch} that the phenomenon 
disappears, i.e., energy equipartition is quickly attained, if
energy is large enough. A long debate  then followed
\cite{izrailevchaos,noichaos,pettinichaos,lichtchaos,bambponno}
 concerning the
questions (still unanswered) whether the phenomenon persists in the 
``thermodynamic limit'' (i.e., when  the number $N$ of particles and the
energy $E$ both grow to infinity with a finite value of the specific
energy $\epsilon=E/N$), and  whether it can be interpreted in a
metastability perspective \cite{fucito, berchialla}. 
Another still open problem is whether
the phenomenon persists when the dimensions are increased, passing
from a chain of particles to a 2-- or a 3--dimensional lattice \cite{benettin}.
\begin{figure*}[ht]
    \includegraphics[width=8.75 truecm]{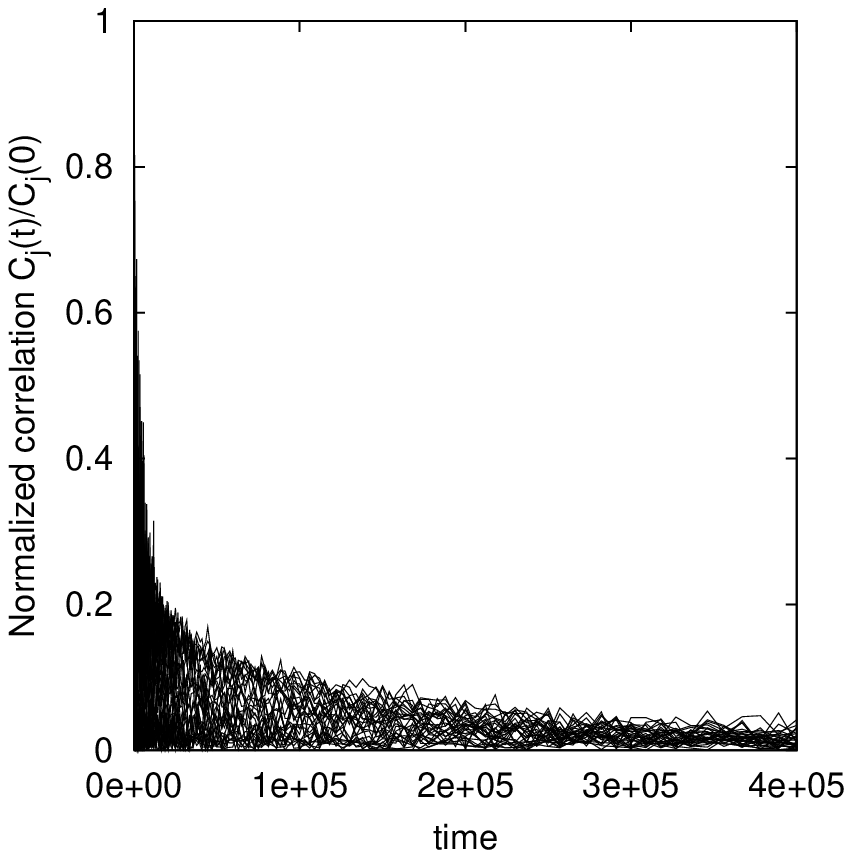} 
    \includegraphics[width=8.75 truecm]{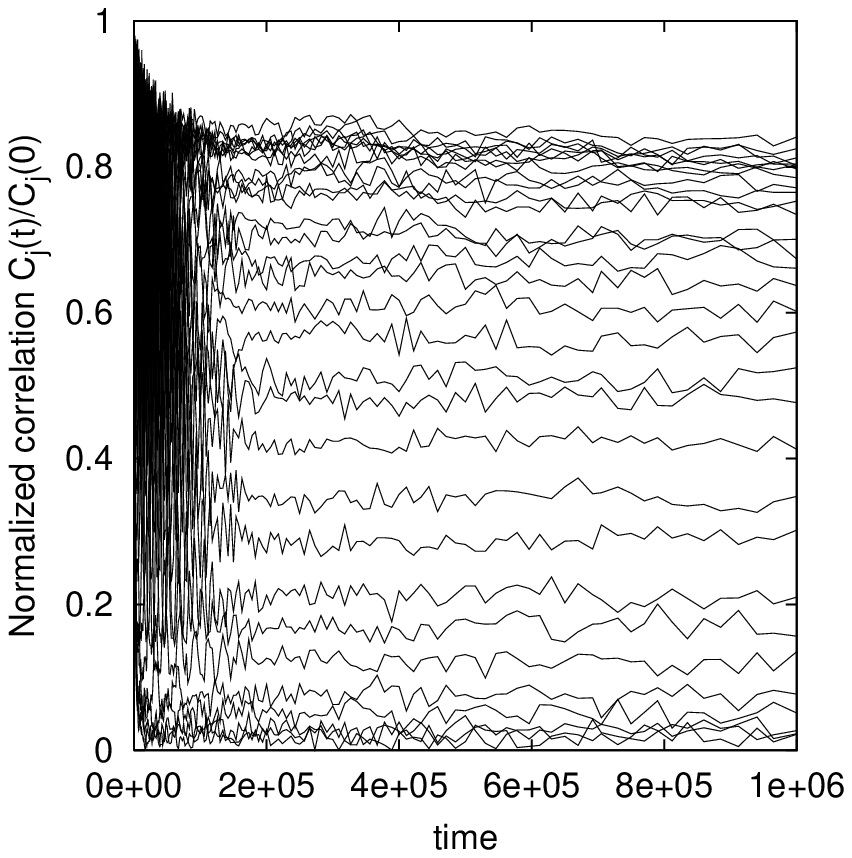} 
 \caption{\label{fig1} The normalized autocorrelation function
   $C_j(t)/C_j(0)$ of the mode--energies $E_j$ versus time, 
for some selected values of
   $j$  ($j=16k$, $k=0,\ldots,32$) and
    $N=511$  at two  values of the specific energy $\epsilon$. Left
   panel, $\epsilon=3.16\ 10^{-2}$, right panel $\epsilon=3.16\ 10^{-3}$.}
\end{figure*}

In the present letter we address a further  problem, namely
whether some analog of the FPU phenomenon occurs 
if generic initial data are taken  rather than some  very special ones
(see also \cite{vulpiani,tenen,noicalore}).
More generally, we would like to look at the problem of the approach
to equilibrium from the viewpoint of ergodic theory, in which   one 
considers  in principle all initial data, weighted with an invariant
measure such as the microcanonical one.  Now, in
ergodic theory it is 
well known that an approach to equilibrium is guaranteed
if a system is proven to be  mixing. Let us recall this. 
For a function $f$ on
phase space, define
$f(t)=f\circ \Phi^t$, where $\Phi^t$ is the
flow induced by a given Hamiltonian. Denote also
by $\langle \ \cdot \ \rangle$  expectation
with respect to the microcanonical  measure. Then, mixing
amounts to 
requiring (see \cite{arnoldavez}, theorem 9.8) that for all  
(square--integrable) functions $f$ and $g$  
the correlation
$$ C(t)= \langle f(t)g(0)\rangle -\langle f(t)\rangle \langle g(0)\rangle
$$
tends to zero as  $t\to\infty$. So,
equilibrium occurs if  the correlations between all pairs of functions
are proven to decay to zero  with increasing time.

The original FPU results, although expressed in terms of
time--averages and observed only for a very special class of initial
data, suggests that, at low energies, the 1--dimensional FPU system
``does not have mixing properties up the considered time''. In the
present letter we give strong numerical indications that this is
actually the case, even in the thermodynamic limit.  
This is obtained by computing the correlations of suitable functions,
the averaging being performed  over initial data sampled out 
from the microcanonical ensemble through
a suitable   Montecarlo method. According to the computations,  
for low enough energies the correlations 
appear to relax to some positive values.  This we call
an FPU--like phenomenon.  Such a phenomenon seems to suggest a
positive property, namely that the system did actually relax to some
well defined anomalous state. We leave for further studies the
questions whether such a result should be interpreted in a
metastability perspective,
and whether it persists for $2$ and $3$--dimensional lattices.

For what concerns the functions to be 
investigated, we  started up  by following  FPU, and    restricted
our  attention to the
normal--mode   energies $E_j(t)$, i.e. we studied 
the autocorrelations
$$
C_j(t)=\langle E_j(t)E_j(0)\rangle - \langle E_j(t)\rangle  
\langle E_j(0)\rangle \ .
$$
It will be shown later, however, that a major role is played by other
related quantities, i.e., the energies $\mathcal{E}_j$ of
``packets'' of nearby modes, whose relevance was pointed out in  
\cite{livikantz} (see also \cite{tenen}).

The aurocorrelations $C_j(t)$ of the mode--energies  $E_j$
were  numerically estimated 
by integrating a sufficiently large number
$K$ of orbits (actually, $K=10000$, apart from two cases which are 
mentioned later), and computing
at any time the arithmetic mean of the values corresponding
to the single orbits. The single initial data were sampled  out
from a microcanonical ensemble  at specific energy $\epsilon $. This was
actually implemented as follows. Each initial datum was extracted
from a Gibbs ensemble  (with the quadratic part only  of the total
Hamiltonian) at temperature $\epsilon$, and was 
then rescaled to let it fit the constraint $H=N\epsilon$ ($H$ being now
the total Hamiltonian).

We took the standard $\alpha$--FPU Hamiltonian, namely, 
\begin{equation*}
    H(p_1,\ldots p_N, x_1,\ldots,x_N)=\sum_{k=1}^N \frac
   {p_k^2}{2m}+\sum_{k=0}^N V(x_{k+1}-x_k)\ , 
\end{equation*}
with $x_0=x_{N+1}=0$,
where $p_k$ is the momentum conjugated to the particle's position
$x_k$, and the interparticle
potential is $V(r)= {r^2}/2 +\alpha {r^3}/3$.
Units were so chosen  that  $m=1$, and $\alpha=1/4$. 
The integrations were performed with the standard leap--frog
(or Verlet) method, with tipical step $0.05$. 

The analog of the FPU phenomenon (together with its disappearing at
high energies) is exhibited in Fig.~\ref{fig1}, where the normalized
autocorrelation functions $C_j(t)/C_j(0)$ of the mode--energies $E_j$
 are plotted versus time, for
some selected values of $j$, with $N=511$ and a sample of $10000$
initial data.  The left and the right panels correspond to a case of a
``high'' specific energy and to a case of a ``low'' specific energy
respectively, precisely, $\epsilon= 3.16\ 10^{-2}$ and $\epsilon=
3.16\ 10^{-3}$. It is seen that in the case of a high energy all
autocorrelations decay to zero essentially within the same
characteristic time, of the order of $10^5$. In the case of a
low energy, instead,  the decay to sero occurs  only for some modes
(with a characteristic time  of
the same order of magnitude as in the previous case), whereas for the 
remaining modes the autocorrelations
appear to have relaxed within that time to some asymptotic
nonvanishing values $c^*(j)$. 

\begin{figure}[t] 
    \includegraphics[width=8.5 truecm]{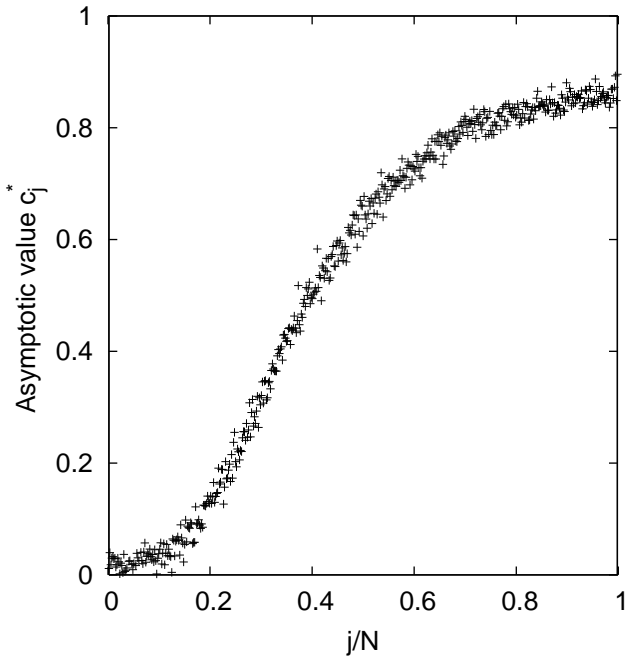} 
    \caption{\label{fig2} The ``asymptotic values'' $c^*(j)$ of the
     normalized  autocorrelation functions $C_j(t)/C_j(0)$ 
versus $j/N$, for the 
      same parameters of Fig.~\ref{fig1}, namely, $N=511$ and
      $\epsilon=3.16\;10^{-3}$. }
  \end{figure}

The natural question then arises of 
understanding whether there is any regularity in the distribution of
the asymptotic values among the modes. We found the interesting
result  that the relevant parameter is the mode--number $j/N$ 
(which is a monotonic
increasing function of the corresponding frequency $\omega_j$).
This is illustrated in Fig.~\ref{fig2}, where the
asymptotic values $c^*$ of the normalized autocorrelations are plotted
versus $j/N$.  The figure refers
to the same values of $\epsilon$, $N$ and $K$ (number of initial data) 
as in the right panel of
Fig.~\ref{fig1}. The first interesting feature is that the data appear
to lie on some smooth curve. Moreover, the shape of the curve shows that
the low--$j$ modes (i.e., the low--frequency ones) are the ones that
exhibit a quick relaxation to the ``final'' expected value $0$, while
the high--frequency modes remain ``frozen''
near   the initial value $1$.  One should notice that it is
precisely by looking at the correlations that the frequency can be
found to play any role, because the microcanonical expectations of the
energies are instead all equal (equipartition).  On the other hand, as
the correlations are well known to play a major role in thermodynamics
according to the fluctuation--dissipation theorem, one may conjecture
that the anomalous behaviour discussed  here might be of physical
interest, for example  for  some  phenomena of anomalous 
decay observed in recent experiments (see \cite{schul}).

  \begin{figure}[t]
    \includegraphics[width=8.5 truecm]{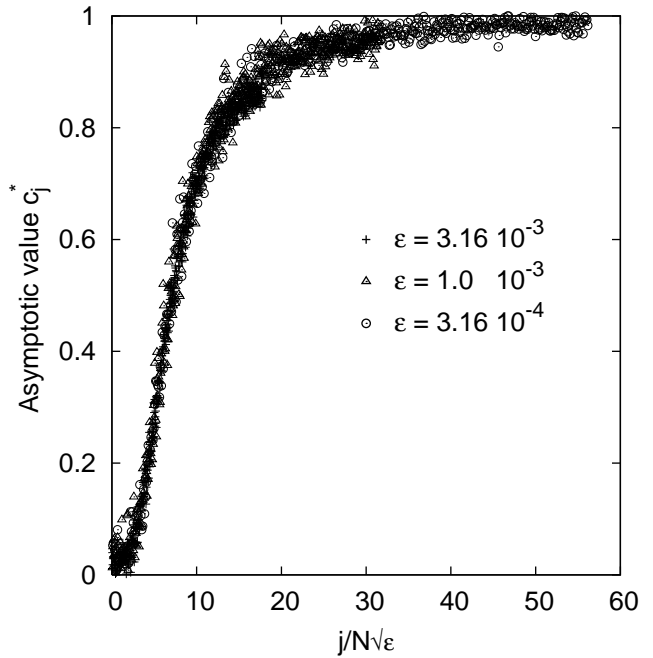} 
    \caption{\label{fig3} The ``asymptotic values'' $c^*(j)$ of
     the  normalized  autocorrelation  functions $C_j(t)/C_j(0)$, 
versus $ \frac
      jN/\sqrt{\epsilon}$,  for $N=511$ and 
      $\epsilon= 3.16\;10^{-3}$,
      $1.0\;10^{-3}$,  $3.16\;10^{-4}$. }
  \end{figure}

We come now to the dependence of the function $c^*(j)$ on the
specific energy $\epsilon$. The curve is expected to reduce to the
straight lines $c^*=0$ and $c^*=1$ for large and small values of
$\epsilon$ respectively. We found the interesting result that, for a
fixed $N$, the curve is a function of only one variable, precisely,
one has $c^*(j,\epsilon)=f(j/\sqrt{\epsilon})$. This is illustrated in
Fig.~\ref{fig3}, where, still for $N=511$, $c^*$ is plotted versus
$(j/N)/\sqrt{\epsilon}$ for three values of $\epsilon$, namely,
$\epsilon= 3.16\;10^{-3}$, $1.0\;10^{-3}$,
$3.16\;10^{-4}$.  
A rather good superposition of the curves seems to
be observed. In particular notice that,  with  increasing $\epsilon$, 
the domain of the curve
shrinks to the left, so that  $c^*$ is found to approach
 the value zero. Thus for large 
$\epsilon$ one has a
complete decay to zero of the correlations, i.e., an analog of the
Izrailev--Chirikov phenomenon.
Notice that for each $\epsilon$ the asymptotic values 
$c^*(j)$ had to be evaluated at a suitable time, 
namely that at which
stabilization just started occurring, in the sense of
Fig.\ref{fig1}. Such a relaxation time was found to increase as
$1/\epsilon$ with decreasing $\epsilon$.
\begin{figure*}[t]
    \includegraphics[width=8.75 truecm]{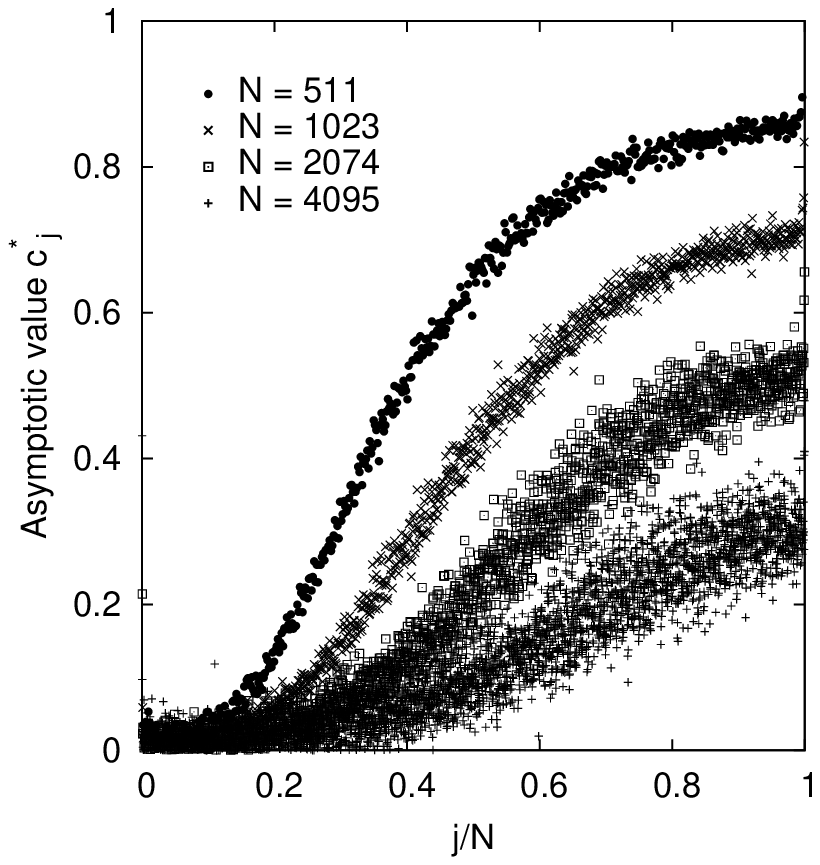} 
    \includegraphics[width=8.75 truecm]{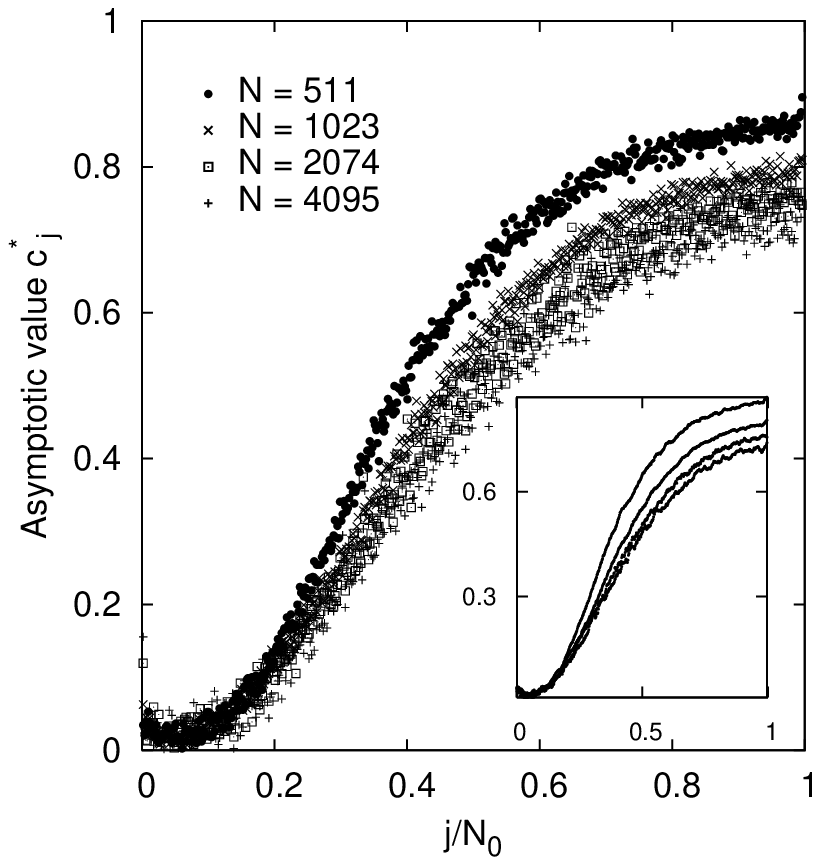} 
 \caption{\label{fig4} The ``asymptotic values'' $c^*(j)$ 
   of the
   normalized autocorrelation functions for the mode--energies $E_j$
   (left panel), and for the energies $\mathcal{E}_j$ of packets of
   $n$ nearby modes with $n$ proportional to $N$ (right panel), for
 an increasing number $N$ of particles. The value of
   $\epsilon$ is the same as
   in Fig.~\ref{fig2}. The asymptotic value $c^*(j)$ is plotted in the left
   panel versus $j/N$ for $j=1,\ldots, N$, and in the right panel 
versus $j/N_0$  for    $j=0,\ldots,N_0-1$, with  $N_0=511$. In the
   inset of the right panel, 
for each of the four series of data the corresponding   
``moving averages'' (over eleven points) are plotted.
}
\end{figure*}

The last point we address concerns the dependence of the results on
the number $N$ of particles. This is a quite 
delicate matter, on which we feel we got an
interesting  result. To begin with we point out that, if one takes a naive
approach  and plots the  curves analogous to that of Fig.\ref{fig2}
 for increasing  values of
$N$, the curves are found to collapse towards  the trivial one
 $c^*=0$.
 This  is shown in Fig.\ref{fig4}, left panel, where the
curves for $N= 511, 1023, 2047, 4095$ are reported, for the same
$\epsilon$ as in Fig.~\ref{fig2}. Concerning 
the number $K$
of initial data, this  had forcedly to be diminished with increasing
$N$, and we had to pass from $K=10000$ for $N=511$ and $1023$ to $K=
5000$ and $2000$ for $N= 2047$ and $4095$ respectively. This, by the
way, explains the  broadening of the  ``curves'' for the  two large
 values of $N$.

It would however be incorrect to infer from such a collapse
that mixing occurs in the thermodynamic limit, because mixing requires
the decaying to zero of the correlations  for all pairs of
functions. Instead, a decay to positive values is observed  if a
suitable choice is made for the functions to be tested for
autocorrelation.
Actually, instead of considering 
the energies $E_j$ of the single modes, we considered
the energies of packets of $n$ nearby modes, with $n$ proportional
to $N$,  precisely,  the $N_0$ quantities
$$
\mathcal{E}_j=\sum_{k=nj+1}^{n(j+1)} E_k\ ,\quad {\rm where}\quad
n=\frac {N+1}{N_0+1}\quad ,\  N_0=511\ ,
$$
for $j=0,1,\ldots,N_0-1$.
In Fig. 4, right panel,  the analog of $c^*$ for the quantities 
$\mathcal{E}_j$ is plotted   versus
$j/N_0$, for the same numbers $N$ and $K$ as in the left panel. 
It is true that the different curves do not superpose, and that a
certain decreasing is observed, especially in passing from $N=511$ to $N=1023$.
But for larger values of $N$ the results seem to indicate
that a  nontrivial limit curve is being  approached. This is better
illustrated in the inset, where, in order to improve the readability
of the graphs, the data were  smoothed out by a standard 
``moving averaging'' with eleven points.  
In our opinion, the results suggest that  the FPU--type
phenomenon discussed here may persist in the
thermodynamic limit, for a one--dimensional chain. And this, not for
very special initial data, but in a global sense involving an averaging
over all initial data, in a microcanonical setting.

It may be of interest in this connection to recall that  analytical 
perturbative estimates for a ``freezing'' in
the thermodynamic limit, much in the same spirit of the present paper, 
i.e, by averaging initial data over an invariant measure, were
obtained very recently (see \cite{caratiJSP}).

\bibliography{biblio_fpu}

\end{document}